\documentclass[aps, pra, reprint]{revtex4-2}
\usepackage{graphicx, color}
\usepackage{amsmath}
\usepackage{mathtools}
\usepackage{subsupscripts,epstopdf}
\usepackage{braket}
\usepackage{hyperref} 
\hypersetup{
    colorlinks=true,
    linkcolor=blue,
    citecolor=blue,
    filecolor=magenta,      
    urlcolor=blue,
    }
\usepackage[utf8]{inputenc}

\newcommand{\bstate}{O$_2^+$~$b^4\Sigma_g^-$}
\newcommand{\astate}{O$_2^+$~$a^4\Pi_u$}
\newcommand{\fstate}{O$_2^+$~$f^4\Pi_g$}
\newcommand{\xstate}{O$_2^+$~$X^2\Pi_g$}

\begin{document}

\title{Rotational coherences in O$_2^+$ following strong-field ionization}



\author{Huynh Van Sa Lam}
\author{Tomthin Nganba Wangjam}
\author{Vinod Kumarappan}
\email[Email: ] {kumarappan@ksu.edu}
\affiliation{James R. Macdonald Laboratory, Kansas State University, Manhattan, KS 66506, USA}


\date{\today}

\begin{abstract}
We investigate the wave packet that remains bound in the ground and excited cationic states of oxygen after strong-field ionization by an intense 800-nm pulse. Much weaker probe pulses (800 or 264 nm) are used to dissociate these still-bound cations. The momentum distribution of O$^+$ is measured as a function of pump-probe delay and Fourier-transformed to obtain kinetic-energy-dependent and rotational-state-resolved quantum beat spectra. The sub-cm$^{-1}$ resolution of the Fourier transform allows unambiguous identification of the electronic, vibrational, and rotational states populated by the pump and then dissociated by the probe. Although strong-field ionization is expected to populate the lower-lying $X^2\Pi_g$ and $a^4\Pi_u$ states more effectively than the $b^4\Sigma^{-}_g$ state, a wave packet in the $X^2\Pi_g$ state is seen only with the 264-nm probe and only weak signatures of the $a^4\Pi_u$ states are found with either probe. The experiment confirms the role of the resonant coupling between the $b^4\Sigma^{-}_g$ and $a^4\Pi_u$ states by the 800 nm pulses [Xue \textit{et al.}, Phys. Rev. A 97, 043409 (2018)] and reveals the importance of rovibrational excitation in determining the momentum distribution of the O$^+$ fragments. The strong $X^2\Pi_g$ state contribution observed with the 264-nm probe also shows the importance of resonant coupling in the probe pulse. The sub-cm$^{-1}$ resolution also resolves spin-orbit splitting in both the $X^2\Pi_g$ and $a^4\Pi_u$ state wave packets.
\end{abstract}

\maketitle

\section{\label{intro}Introduction}
When molecules are ionized by intense ultrashort laser pulses, a range of electronic, vibrational, and rotational states can be coherently populated in the cation either directly in the ionization step itself, or through efficient dipole coupling between various cationic states after ionization (see, for instance, \cite{De2010,Yao2016,Lam2025PRAL}). The resulting dynamics of these wavepackets contain information not only about the ionization and excitation processes in the ionizing laser pulse, but also about any subsequent coupling between different degrees of freedom.   

The dissociation of $\rm{O}_2^+$ by ultrafast laser pulses has been investigated under various conditions starting either from the neutral molecule or an ion beam. De \textit{et al.} \cite{De2010} ionized neutral molecules with a few-cycle 800-nm pump pulse and then dissociated the ions with a similar probe pulse. Corlin \textit{et al.} \cite{Corlin2015} and Malakar \textit{et al.} \cite{Malakar2018} used XUV pulses to ionize the molecules and an 800-nm probe pulse to dissociate them. Zohrabi \textit{et al.} \cite{Zohrabi2011} use a $\rm{O}_2^+$ ion-beam target and a single 800-nm pulse for dissociation. In all these experiments, the probe pulse did not have enough intensity to dissociate the \xstate ~ground state and the primary mechanism was determined to be the coupling between the $a{}^4\Pi_u$ ~state and the weakly-repulsive and metastable \fstate ~state \cite{Zohrabi2011, Magrakvelidze2012, Corlin2015, Malakar2018}.   

However, in the IR-pump--IR-probe experiment by De \textit{et al.} \cite{De2011}---the only aforementioned experiment employing strong-field ionization to launch the cationic wave packet---there are notable discrepancies between the measured and simulated kinetic energy release (KER) and quantum beat (QB) spectra. Most importantly, the simulated energy-dependent structure in the KER and the QB frequencies do not match the measurement. Xue \textit{et al.} \cite{Xue2018} have theoretically reexamined this experiment and identified the contribution of the $b{}^4\Sigma_g^-$ state to the dissociation of $\rm{O}_2^+$. Although this state is not dipole-coupled to the repulsive $f{}^4\Pi_g$ state, it is resonantly coupled to the $a{}^4\Pi_u$ state by a 800-nm photon. By taking into consideration the population redistribution between the $a{}^4\Pi_u$ and $b{}^4\Sigma_g^-$ states, Xue \textit{et al.} can reproduce the energy-dependent structure and the quantum beat frequencies observed in \cite{De2011}, illustrating the importance of post-ionization population redistribution in understanding the dissociation dynamics.

In this work, we use strong 800-nm, 30--45-fs pulses to ionize $\rm{O}_2$ and then use weak pulses (either 800 nm or 264 nm) to probe the wave packet that remains bound in cationic states. We measure the delay dependence of the momentum spectrum of $\rm{O}^+$ ions over multiple rotational periods and generate high-resolution KER-dependent QB spectrum using Fourier transformation \cite{Forbes2018}. Our results show a dominant contribution from rotational states of $b{}^4\Sigma_g^-$ with both probe wavelengths, corroborating the theoretical work by Xue \textit{et al.} \cite{Xue2018}. We also show the importance of rovibrational excitation in determining the momentum spectrum of O$^+$ fragments. The characteristics of the $X^2\Pi_g$ wavepacket observed with the 264-nm probe pulse highlights the importance of resonant coupling during the probe pulse and the role of spin-orbit coupling in the rotational dynamics of the cation.

\section{\label{sec:med} Methodology}    

The experimental setup has been reported elsewhere \cite{Lam2022, Lam2020, Wangjam2021} and is described only briefly here. We split the output of a Ti:Sapphire laser into pump and probe arms, introduce a delay between the two, and recombine them on a dielectric beamsplitter. The probe pulse is either the IR pulse or its third harmonic generated using BBO crystals. Both are then focused into a velocity map imaging (VMI) spectrometer \cite{Eppink1997} by a 20-cm-focal-length dielectric mirror coated for both wavelengths. All pulses have identical linear polarization, parallel to the detector plane. The pump has enough intensity ($\approx3\times10^{14}$~W/cm$^2$) to dissociate $\rm{O}_2$ into $\rm{O}^+$, mainly from singly charged molecular ions; the smaller signal from doubly charged ions is not of interest in this article. The probe is weaker ($2\times10^{13}$ W/cm$^2$ for 800 nm and $1.5\times10^{12}$ W/cm$^2$ for 264 nm) and does not produce any O$^+$ signal. The laser pulses interact with rotationally cold molecules (estimated temperature is $\approx$~2~K \cite{Wangjam2021}) produced by supersonic expansion (0.5\% $\rm{O}_2$ in He at a total pressure of 900 psi) through an Even-Lavie valve \cite{Hillenkamp2003}. The microchannel plate detector in the VMI spectrometer is gated to select O$^+$ ions, and the momentum distribution is recorded by imaging the phosphor screen with a 1000-fps camera and counting individual ion hits using a centroid-finding algorithm. The count rate of $\rm{O}^+$ ions is $<5$ hits per shot to minimize detector saturation at a low kinetic energy (KE). We vary the delay between the pump and the probe using a computer-controlled translation stage. In about 40 hours, we record 15 scans in total; each scan is 200 ps long with a step size of 50 fs (4000 delay points).

We synchronize a mechanical chopper (250 Hz) in the probe beam, the pulsed Even-Lavie valve (500 Hz), camera trigger (1 kHz) and the laser pulses, so that we can classify each laser shot into one of four types: pump-probe-gas, pump-probe-background, pump-gas, and pump-background. We use this information for data normalization, where the corrected image at each delay is
\begin{equation}
M_{2D}(\vec{k}_{2D},t) = \frac{{\rm{Im}\left[ {{\rm{pump, probe, gas}}} \right] - \rm{Im}\left[ {{\rm{pump, probe}}} \right]}}{\rm{Yield\left[pump, gas\right] - Yield\left[pump\right]}},
\label{eq:signal}
\end{equation}
where $\vec{k}_{2D}$ is the projected momentum of $\rm{O}^+$ ions on the detector plane, Im[pump, probe, gas] and Im[pump, probe] are 2D VMI images, and Yield[pump, gas] and Yield[pump] are the total yields of $\rm{O}^+$ ions corresponding to different configurations of pump, probe, and gas. $t$ is the pump-probe delay.

We use the pBasex algorithm \cite{Garcia2004} to invert these corrected 2D projections into 3D distributions of $\rm{O}^+$ momentum in the laboratory frame. These 3D distributions are written as
\begin{equation}
    M(\vec{k},t) = \sum_{k_0L}C_{k_0L}(t)e^{-\frac{(k-k_0)^2}{2\sigma}}P_L(\cos{\theta}),
\end{equation}
where $k$ is the momentum, $\sigma$ is the width of the Gaussian basis, and $P_L(\cos{\theta})$ is the Legendre polynomial of order $L$. The expansion is restricted to small $L$s for small $k$. The detailed data analysis method, including normalization, error propagation, and Abel inversion, has been reported in \cite{Lam2022, Lam2020}.

Subsequently, we transform momentum to KER and calculate the power spectrum of $C_L(\mathrm{KER}, t)$ (QB spectra) by performing FFT along the delay axis. The angle-integrated yield is obtained from the $L=0$ moment while the $L=2$ moment, which depends linearly on $\cos^2\theta$, characterizes the alignment of the momentum distributions with the laser polarization. FFT spectra show quantum beat frequencies where lines correspond to energy differences between two states with a definite phase difference. Finally, a peak detection algorithm was used to locate the peaks in the FFT spectrum.

We use molecular constants from spectroscopic data to calculate rotational energy levels and potential QB frequencies (see Appendix~\ref{sec:Dunham_expansion}). By comparing experimental and calculated QB frequencies, we identify the rotational, vibrational, and electronic states involved in the cationic wave packet. This data is provided in Appendix~\ref{sec:QB_freq}. A comparison of measured KER distributions with the expected values for single- and few-photon dissociation also allows us to identify potential dissociation pathways.
\begin{figure}[h]
\begin{minipage}[h]{\columnwidth}
\includegraphics[width=\textwidth]{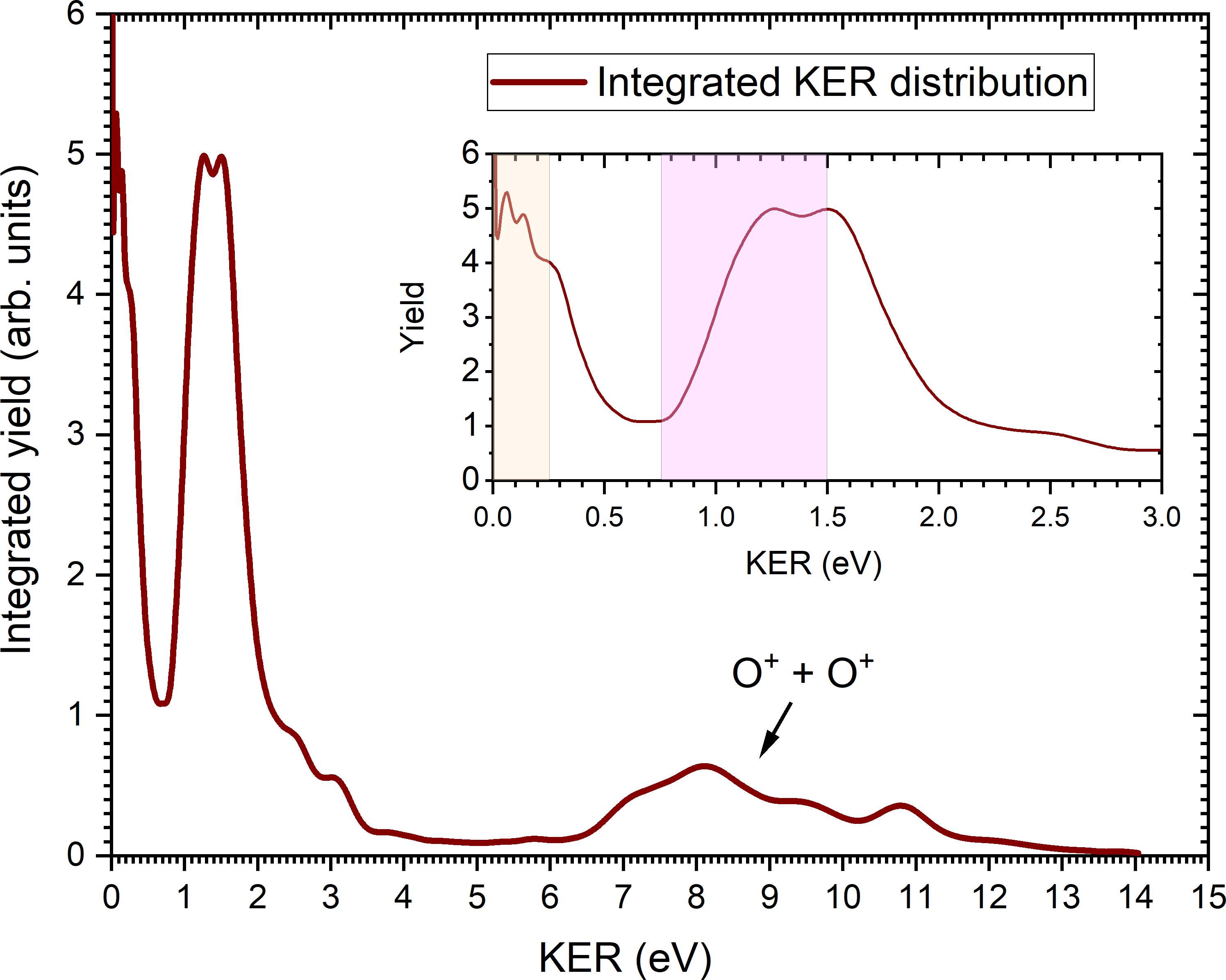}
\caption{Angle-integrated KER spectrum summed over all delays from 1 to 198.95 ps. The 0--250 meV and 0.75--1.5 eV regions, where FFT spectra show bound wave packets, are highlighted by colored vertical bands in the inset.}
\label{fig:Int_KER}
\end{minipage}
\end{figure}

\section{\label{sec:results} RESULTS AND DISCUSSION: IR PROBE}    
\subsection{\label{sec:KER spectrum} Kinetic Energy Release}

\begin{figure*}
\begin{minipage}[h]{\textwidth}
\includegraphics[width=\textwidth]{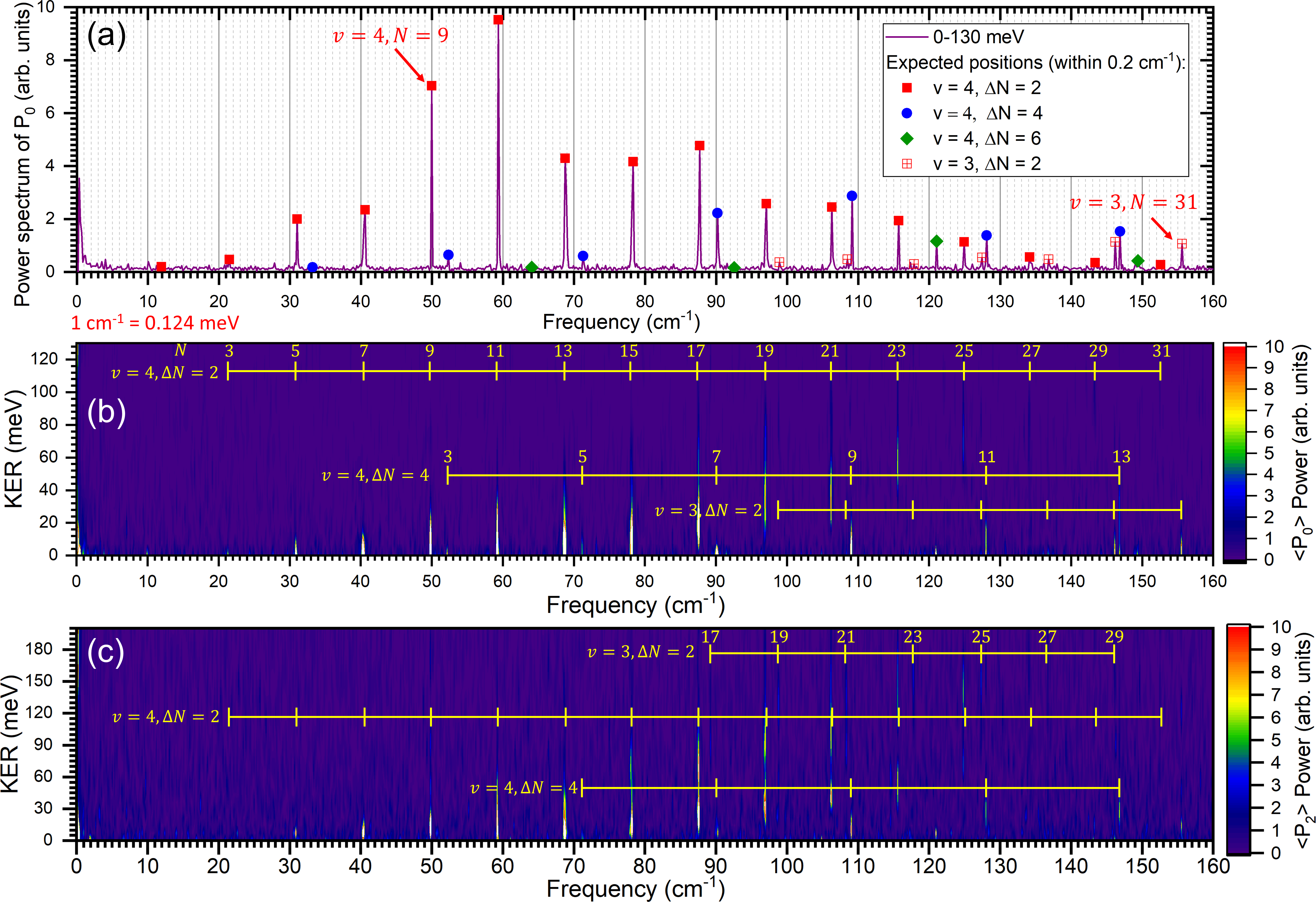}
\caption{Quantum beat spectra of the 0-250 meV band. (a) Power spectrum of O$^+$ yield ($=\braket{P_0}$), with the KER integrated from 0 to 130 meV. Calculated QB frequencies are shown as scatter plots, where each point is shown at the calculated frequency along the horizontal axis and displaced vertically to the top of the corresponding peak in the FFT spectrum for visual clarity. We do not calculate the intensities of these peaks. (b) The KER-dependence of the power spectrum. (c) Power spectrum of $\braket{\cos^2\theta}$ ($\propto\braket{P_2}$). Please note the larger KER range. In panels (b) and (c), the calculated rotational QB frequencies are shown as horizontal scales where $\nu$ is the vibrational quantum number, $N$ is the rotational quantum number of the lower rotational state, and $N+\Delta N$ is the upper state. All quantum numbers are for $b{}^4\Sigma_g^-$.}
\label{fig:0_250_meV}
\end{minipage}
\end{figure*}

In this Section, we discuss experiments in which the probe pulse is $\approx$ 800 nm, 45 fs, $2\times10^{13}$ W/cm$^2$ and the pump pulse is $\approx$ 800 nm, 40~fs, $3\times10^{14}$ W/cm$^2$. Figure~\ref{fig:Int_KER} shows the KER spectrum integrated over all delays from 1 to 198.95 ps; the pump-probe overlap region was excluded. The energy calibration is in agreement with previous experiments \cite{De2011, Malakar2018, Yu2020}. The high-energy $\rm{O}^+$ ions from the O$^{2+}_2\longrightarrow$ O$^++$O$^+$ channel (KER $>$ 5 eV) do not show quantum beats in the FFT spectrum, indicating that they originate from the interaction with the pump alone. This channel is not of interest here and will not be discussed further in this article. Instead, we focus our attention on two bands in the low-energy region --- 0--250 meV and the 0.75--1.5 eV --- where FFT spectra show bound wavepackets. These regions are highlighted by colored bands in the inset of Fig.~\ref{fig:Int_KER}.

\subsection{\label{sec:0_250meV} The 0-250 meV band}  

\begin{figure}[h]
\begin{minipage}[h]{\columnwidth}
\includegraphics[width=0.975\textwidth]{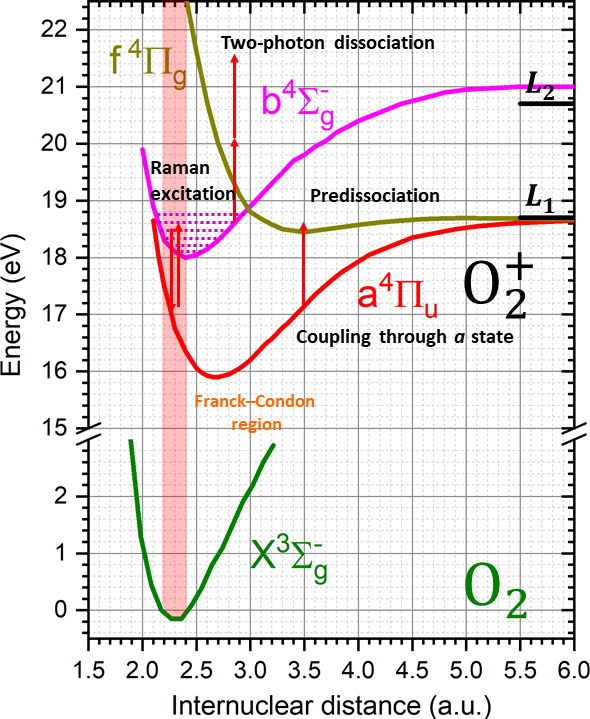}
\caption{Sketch of relevant potential energy curves (adapted from Ref.~\cite{Krupenie1972} and \cite{Xue2018}) for the dissociation of the $b^4\Sigma^{-}_g$ state. The pump excites a broad rotational wave packet in \bstate{}. This bound wave packet can then predissociate or dissociate by the probe via multiple pathways as discussed in Sec.~\ref{sec:results}.}
\label{fig:PES_b}
\end{minipage}
\end{figure}

The QB spectra of the 0-250 meV band are shown in Fig.~\ref{fig:0_250_meV}, where panel (a) shows an FFT spectrum of the delay-dependent yield integrated over the range of KER from 0 to 130 meV. These spectra show that the delay dependence is dominated by $\nu=4$ of the O$_2^+$ $b^4\Sigma^-_g$ state, but weaker contributions from $\nu =3$ can also be identified. Importantly, in this KER range, there are no peaks that cannot be attributed to these two vibrational states. This is in contrast with the assignment made by De \textit{et al.}~\cite{De2011} in an experiment using 8-fs pump and probe pulses, where this energy range was assigned to vibrational levels of the O$_2^+$~$a^4\Pi_u$  state. As pointed out by Xue \textit{et al.}~\cite{Xue2018} and confirmed by the excellent agreement between the observed and calculated rotational QB frequencies in our experiment, low-energy oxygen ions result from the dissociation of O$_2^+$ in the $b^4\Sigma^-_g$ state, not in the $a^4\Pi_u$ state. Please refer to Fig.~\ref{fig:PES_b} for a sketch of relevant potential energy curves.

\begin{figure*}
\begin{minipage}[b]{\textwidth}
\includegraphics[width=\textwidth]{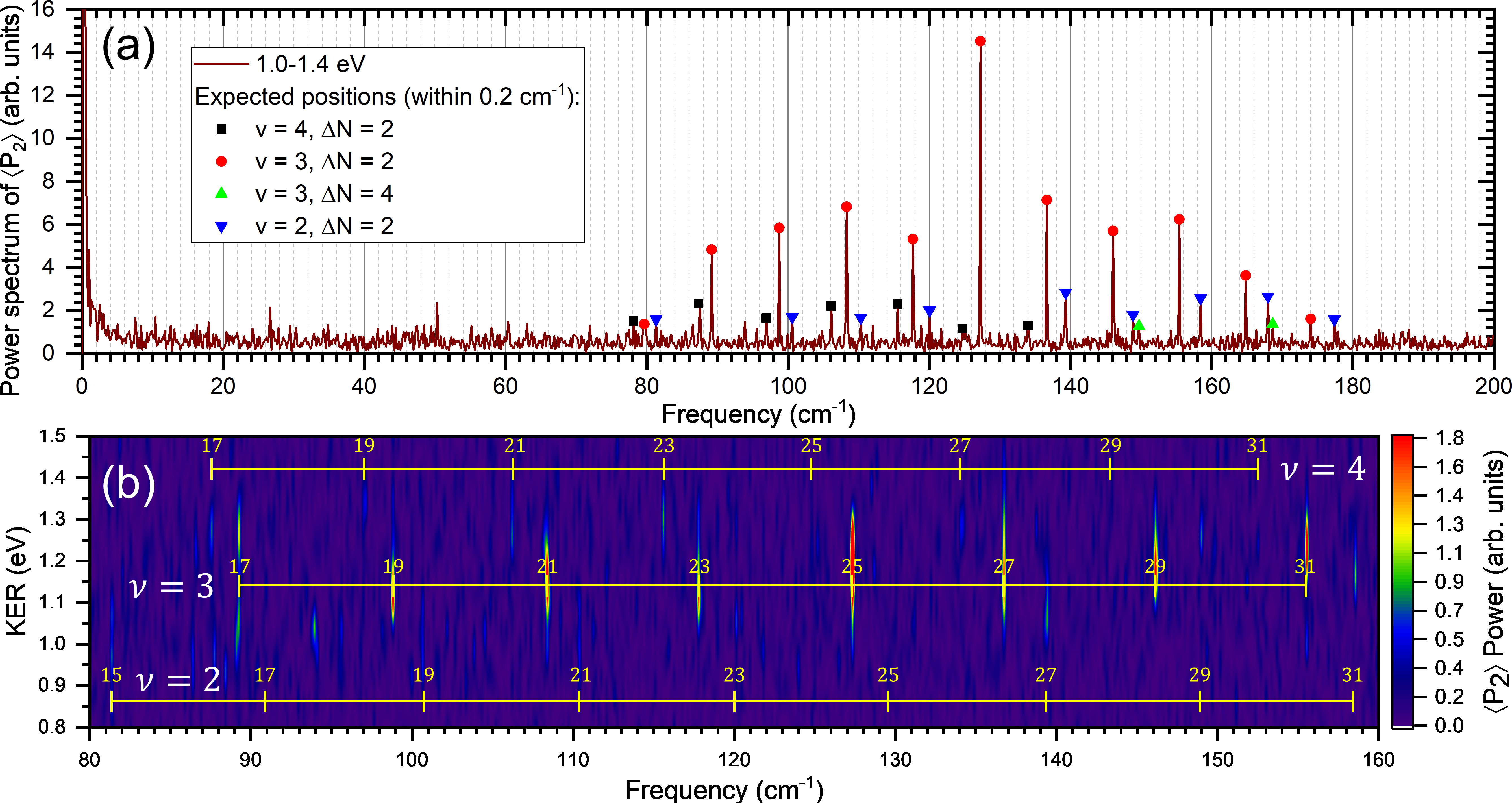}
\caption{FFT spectra of the delay-dependent KER for $\braket{P_2}$ of the high energy band (0.75-1.5 eV). In panel (a), the KER is integrated over the 1.0 -- 1.4 eV range. In panel (b), the KER is not integrated. The dominant channel is $\nu=3$, channels $\nu=4$ and $\nu=2$ are weaker. In this case, the $b^4\Sigma^-_g$ state absorbs two photons and dissociate into the second dissociation limit. The calculated Raman frequencies are shown for comparison. In panel (a), these frequencies are plotted at the height of the experimental peaks for clarity; however, we do not calculate the intensity of the Raman peaks.}
\label{fig:750_1500_meV}
\end{minipage}
\end{figure*}

The first dissociation limit ($L_1$) is just $\approx$~2~meV below $N=9$ for $\nu=4$ and $N=31$ for $\nu=3$ \cite{Tadjeddine1978} as marked in Fig.~\ref{fig:0_250_meV}(a), respectively. Predissociation of rotational levels $N \ge 9$ can give O$^+$ ions with small KER in this range (determined by the energy molecules have in excess of the dissociation limit). Indeed, $b{}^4\Sigma_g^-$ can predissociate via spin-orbit coupling with the $f^4\Pi_g$ and $d^4\Sigma_g^+$ states~
\cite{Tadjeddine1978}. However, any predissociation not influenced by the probe pulse will appear as a peak at zero frequency in the KER-FFT spectrum and cannot be discerned. In the FFT spectrum away from zero frequency, the yield, the KER, or the angular distribution of the O$^+$ ions must be modulated by the probe pulse.

The data shows a strong rotational excitation --- a broad rotational wave packet up to $N=31$ --- by the pump (note that the neutral ground state is predominantly $N=1$ at $\approx$ 2~K \cite{Wangjam2021}). In the pump pulse, molecules absorb energy via multiple rotational Raman cycles --- either in the neutral (pre-ionization) or cationic \astate{} and \bstate{} states (post-ionization) --- and ascend well past $N=9$, the threshold for predissociation from $b{}^4\Sigma_g^- (\nu=4)$. Nevertheless, it is possible to observe these states in our FFT data because the predissociation lifetimes of $v = 4, N>7$ lie between 300 and 800~ps (depending on fine structure level~\cite{Moseley1979}), significantly longer than our 200 ps pump-probe scan. Predissociation will broaden the peaks in the FFT spectrum --- the full width at half-maximum of the resulting Lorentzian line profile would be $(c\tau)^{-1}$, where the speed of light $c$ is in ps/cm and the lifetime $\tau$ is in ps. Since the FFT bin size is determined by the same expression but with $\tau$ representing the duration of the scan ($\approx200$ ~ps), line-broadening has little effect on these lines in our experiment. Weaker contributions from $\nu=3$ are also seen in Fig.~\ref{fig:0_250_meV}. Here, too, the signal is strongest near the predissociation threshold ($N=31$). In this energy range, there are no discernible contributions from any other vibrational states.       

Note that while the presence of any particular QB frequency in the KER-FFT spectra indicates a pump-excited coherence, the kinetic energy distribution is determined by both the initial states and the probe process. Hence, the KER distribution helps to clarify the dissociation pathways in the probe pulse. The near-zero KER in Fig.~\ref{fig:0_250_meV}(b-c) suggests net-zero-photon dissociation of the $\nu=4$ state. As expected, the KER increases with increasing vibrational and rotational quantum numbers. In addition, there are telltale signatures of energy gain from the probe: lines corresponding to states that are below the predissociation threshold are present, the KER distribution has significant vertical spread, and some $v=3$ lines exhibit higher KER than $v=4$ lines.    

The data suggest a two-part mechanism for the dissociation of O$_2^+$ $b{}^4\Sigma_g^-$, both mediated by Raman processes. First, rotational and vibrational Raman excitation raises the population from below the predissociation threshold to within about 100 meV (the laser bandwidth). The final two-photon transition transfers the population to the $f^4\Pi_g$ state, leading to dissociation. According to Xue \textit{et al.}~\cite{Xue2018}, these processes are mediated by the $a^4\Pi_u$ state. However, because of the proximity of the crossing of the $b{}^4\Sigma_g^-$ and $f^4\Pi_g$ states to the outer turning point of vibrational motion in the former, nonresonant two-photon (or electronic Raman) transitions may also directly transfer population to the continuum of the latter. This alternative mechanism, which may be viewed as laser-induced predissociation \cite{Lau1977, Lau1978} or in terms of a two-photon light-induced conical intersection \cite{Kuebel2020}, does not require resonant transitions to an intermediate state. The presence of near-zero KER lines from $v=4$ in the data obtained with a 264~nm probe (as discussed in Sec.~\ref{sec:UV}) indicates that this nonresonant mechanism plays a significant role in the dissociation of the $b{}^4\Sigma_g^-$ state.

Vibrational states with $v \ge 5$ have no energy barrier to predissociation. The coupling to the $f^4\Pi_g$ state in the tail of the pump pulse is likely to be efficient for these states, leaving behind little or no population for the probe to dissociate. On the other hand, $v \le 2$ may be too far below the threshold to be reached by our weak probe pulse. Thus, the absence of these vibrational levels in our data does not necessarily imply that the pump did not populate them.     

Only weak contributions from the $a{}^4\Pi_u$ state can be found in the spectrum. The data shows a cluster of frequencies around 0.8 eV KER, several lines can be assigned to the $a$ state, but the signal intensity is low and the signal very noisy and will not be discussed in detail. The likely reason for the weak signal is that the pump efficiently dissociates ($v\ge9$ require only a single photon)  or excites to \bstate{} all the vibrational levels that the weaker probe could have reached. The model by Xue \textit{et al.} \cite{Xue2018} has already shown a significant population redistribution from the $a$ state to the $b$ state with a 15 fs pulse. Our pump pulse is longer (40 fs), and its intensity ($\approx 3 \times 10^{14}$ W/cm$^2$) is also higher than the saturation intensity of the $\rm{O^+}$ component, estimated to be $\approx 2 \times 10^{14}$ W/cm$^2$ for a 100 fs pulse \cite{Hishikawa2001}. With a 264 nm probe, discussed in Sec.~\ref{sec:UV}, a clear contribution of low vibrational levels in \astate{} can be identified.     

\subsection{\label{sec:750_1500meV} The 0.75-1.5 eV band}     
Figure~\ref{fig:750_1500_meV} shows FFT spectra of the delay-dependent KER for $\braket{P_2}$ of the high-energy band (0.75-1.5 eV). In panel (a), the KER is integrated over the range of 1.0 to 1.4 eV. In panel (b), the KER is not integrated. We can identify frequencies corresponding to $\nu=2$, 3 and 4 of the $b$ state, with $v=3$ dominant. As expected, the KER increases with $\nu$ while the QB frequency decreases. The KER for $v=4$ is $\approx1.3$~eV. Surprisingly, there is only one clearly identifiable peak (at 127.3~cm$^{-1}$) in the FFT of the ion yield ($\braket{P_0}$). This feature is not yet understood.

\begin{figure*}
\begin{minipage}[b]{\textwidth}
\includegraphics[width=\textwidth]{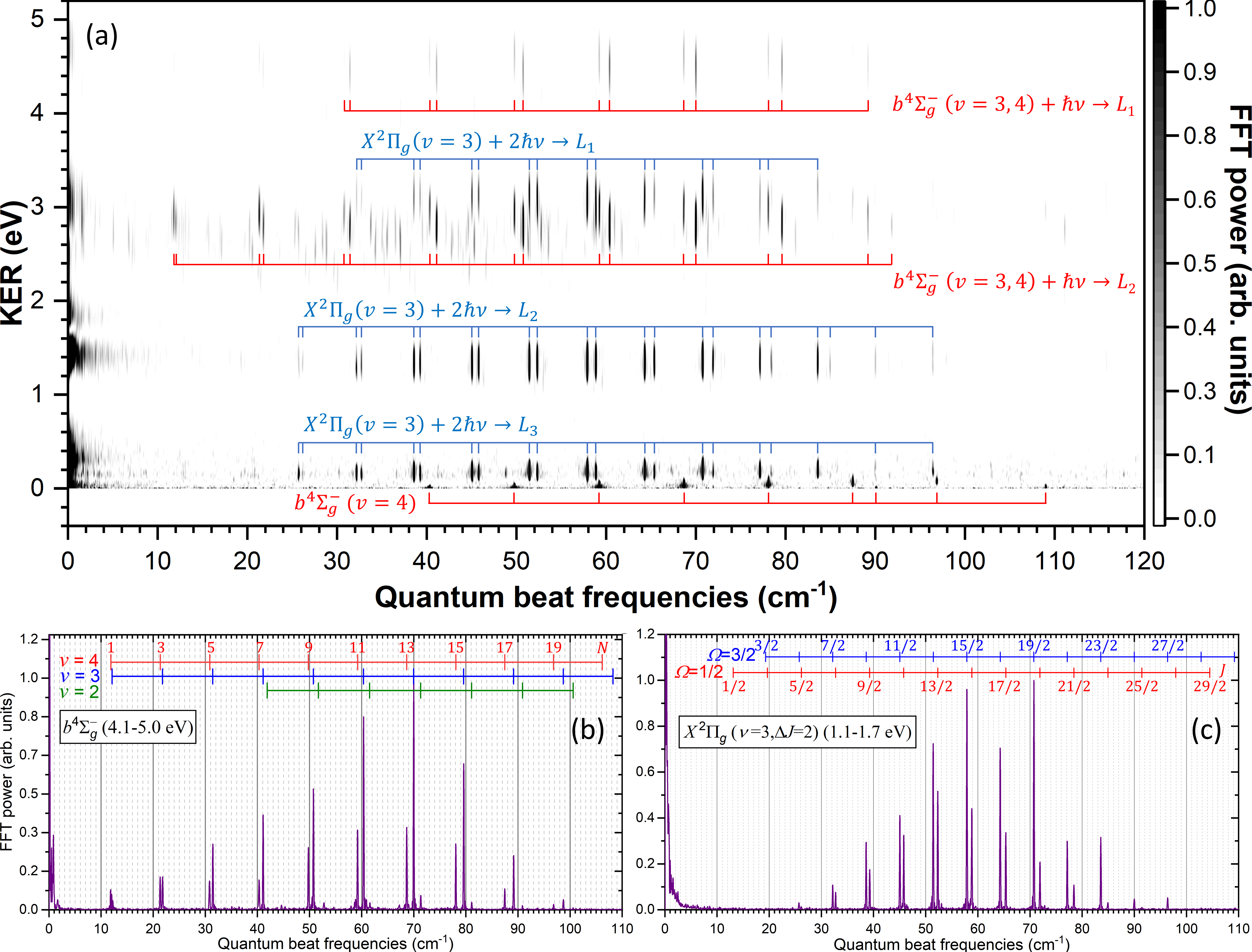}
\caption{(a) FFT spectra of the delay-dependent KER for $\braket{P_0}$ in the IR-UV pump-probe experiment. The color scale is intentionally saturated to show channels of weaker intensity. Panel (b) shows the FFT spectrum of the $b^4\Sigma^{-}_g$ state dissociated into the first dissociation limit ($L_1$) after absorbing one UV photon (the KER is integrated over the range of 4.1 to 5.0 eV). Panel (c) shows the FFT spectrum of the $X^2\Pi_g$ state dissociated into the second dissociation limit ($L_2$) after absorbing two UV photons. The KER is integrated over the range of 1.1 to 1.7 eV. The expected positions of rotational Raman lines corresponding to $\Delta N = 2$ (b) and $\Delta J = 2$ (c) are shown above the spectra, $\Delta N = 4$ and $\Delta J = 4$ are not labeled. For $b^4\Sigma^{-}_g$, only odd $N$ is allowed, while for $X^2\Pi_g$, $N$ is not a good quantum number due to spin-orbit coupling. See Sec.~\ref{sec:UV} for more details.}
\label{fig:UV_probe}
\end{minipage}
\end{figure*}

With 800 nm (1.55 eV photons), two-photon absorption from $b{}^4\Sigma_g^- (v=4)$ followed by dissociation to the second dissociation limit ($L_2$) would result in slightly above 1.1 eV KER \cite{Marian1982} since $L_2$ is 1.97~eV above $L_1$. Adding rotational energy and potentially excitation to $v>4$ in the probe pulse before excitation to the dissociating curve would account for the somewhat larger observed KER. $\nu=3$ and lower will have correspondingly lower KER. This is likely to be the pathway responsible for the dissociation channel seen in Fig.~\ref{fig:750_1500_meV}. Apart from $b{}^4\Sigma_g^-$ itself, there are five states (all are quartets: ${1}^4\Sigma_u^-$, ${2}^4\Pi_g$, ${1}^4\Delta_g$, ${2}^4\Pi_u$, ${1}^4\Delta_u$) that correlate to the $L_2$ asymptote \cite{Liu2015}. Three of them (${1}^4\Sigma_u^-$, ${2}^4\Pi_u$, ${1}^4\Delta_u$) are ungerade and cannot be reached by two-photon transitions from $b{}^4\Sigma_g^-$. ${2}^4\Pi_g$ is energetically ruled out (vertical energy from the right turning point of $b(v=4)$ to the ${2}^4\Pi_g$ state is more than 5~eV, see Fig. 2 of \cite{Liu2015}). Therefore, only the $1 ^4\Delta_g$  remains as a candidate. Thus, we can uniquely identify the intermediate state through which the ion dissociates in this energy range.

Xue \textit{et al.} discussed another possible mechanism that can produce ions in this band (1.3--2.0~eV in \cite{De2011}): predissociation of the B${}^2\Sigma_g^-$ state \cite{Evans1999}, which has a vibrational spacing similar to ${b}^4\Sigma_g^-$. The rotational constants of these two states are sufficiently distinct that we should be able to identify contributions from B${}^2\Sigma_g^-$ in our FFT spectra. In our data we cannot identify any lines that match the rotational levels of B${}^2\Sigma_g^-$. However, the predissociation lifetimes of $v=0$--7 levels are variously estimated to be less than 2~ps \cite{Evans1999} to approximately 70~ns \cite{Richard‐Viard1985}. If the former is correct, the lifetime is too short for our measurements. Thus, we cannot conclusively rule out the population of B${}^2\Sigma_g^-$ in the experiment. Dissociation through $1 ^4\Delta_g$ to the second asymptotic is a simpler explanation for KER is this energy range.     

\section{\label{sec:UV}RESULTS AND DISCUSSION: UV PROBE}

\subsection{Multiple states can be observed in the KER-FFT spectrum with the third harmonic probe}
In this experiment, we use the third harmonic of the fundamental ($\approx$~264~nm, 1.5~TW/cm$^2$, 120~fs), generated by using BBO crystals, as the probe pulse. There is no resonant coupling between $a^4\Pi_u$ and $b^4\Sigma^{-}_g$ states at this probe wavelength. The intensity of this UV probe is kept low such that no ionization of O$_2$ is observed with the probe alone. The bandwidth of the UV probe is about 40 meV ($\approx$ 2.4 nm), narrower than the bandwidth of the 800 nm probe ($\approx$~120 meV) in the previous section and the vibrational spacings ($\approx$~100~meV). Hence, rovibrational excitation by the UV probe is less likely to occur.

The pump pulse ($\approx~3\times$10$^{14}$ W/cm$^2$, $30$ fs, 800 nm) used in this UV-probe measurement has a similar intensity but a shorter duration compared to the pump used in the IR-probe measurement. With lower fluence in the pump pulse, we have less rotational excitation, resulting in a narrower rotational wave packet and quantum beat spectra with lower maximum frequencies. The experiment was conducted using the same step size and delay range as in the previous section, giving an identical frequency resolution in FFT spectra. The dissociation of $b^4\Sigma^{-}_g$ and $X^2\Pi_g$ states into multiple asymptotes was observed in the KER-FFT spectrum in Fig.~\ref{fig:UV_probe} and is discussed in detail below. We also see a weaker signal from quantum beats in the $a$ state, but these channels are not of interest to the present study.

\subsection{\label{sssec:b}$b^4\Sigma^{-}_g(v=2-4)$ state}
In the UV probe, molecules in $b^4\Sigma^{-}_g$ can absorb one photon and dissociate into the first dissociation limit $L_1$ ($\mathrm{KER\approx4.6~eV}$) and the second dissociation limit $L_2$ ($\mathrm{KER\approx2.9~eV}$) (please see Fig.~\ref{fig:PES_b} for potential curves). These channels can be seen in Fig.~\ref{fig:UV_probe}(a). In Fig.~\ref{fig:UV_probe}(b), we show the FFT spectrum of the yield for dissociation of $b^4\Sigma^{-}_g$ into $L_1$ since it is energetically separated from other channels. The frequencies are in excellent agreement with the expected Raman lines; see Table.~\ref{tab:b_800nm_onephoton} for a comparison. Dominant frequencies are assigned to rotational quantum beats of $b^4\Sigma^{-}_g(v=3,4)$ with a much weaker signal from $\nu=2$.

From potential energy curves calculated by Liu \textit{et al.} \cite{Liu2015}, only the $a^4\Pi_u$ state appears as a potential candidate for the dissociation pathway of $b^4\Sigma^{-}_g$ into $L_1$ [O$^+({}^4S)+$ O$({}^3P)$, the lowest energy asymptote] following strict dipole selection rules. $c^4\Sigma^{-}_u$ and $2^4\Pi_u$ satisfy the selection rules to be potential candidates for dissociation of $b^4\Sigma^{-}_g$ into the second asymptote $L_2$ [O$^+({}^4S)+$ O$({}^1D^0)$]. Please refer to Table.~\ref{tab:b_asymptote} in Appendix~\ref{sec:states_n_limits} for a summary of states and their dissociation limits.

The net-zero photon channel of the $b^4\Sigma^{-}_g(v=4)$ state is also observed near zero KER. With the 800 nm probe in Sec.~\ref{sec:results}, we could not distinguish between resonant two-photon coupling between $b^4\Sigma^{-}_g$ and $f^4\Pi_g$ via $a^4\Pi_u$ (at the turning points of the vibrational motion in $a^4\Pi_u$) and nonresonant coupling between them at the curve crossing. However, as mentioned before, there is no resonant coupling between $a^4\Pi_u$ and $b^4\Sigma^{-}_g$ states at this probe wavelength; the probe bandwidth is also narrower than the vibrational spacing. Therefore, only nonresonant coupling via purely Raman excitation plays the role here. Molecules in the $b^4\Sigma^{-}_g(v=4)$ state are further ``aligned'' by the probe pulse, going up the rotational ladder to near the dissociation limit, and then a two-photon coupling to $f^4\Pi_g$ leads to dissociation. Unlike Sec.~\ref{sec:results}, we do not observe any significant contribution from $v=3$ in this energy range. As explained earlier, our rotational wave packet is now narrower due to lower pump fluence and the rotational states on $v=3$ are well below the dissociation limit ($N_{max}<31$). The UV probe pulse with much lower fluence and narrower bandwidth cannot provide enough excess energy via purely rotational or rovibrational Raman excitation for $v=3$ to dissociate in this net-zero photon regime.

The appearance of $b^4\Sigma^{-}_g$ $v=2$, 3, and 4 in measurements using both IR and third-harmonic probes at higher KER suggests that these levels are efficiently populated by the pump and play an important role in the strong-field-induced dynamics in O$_2^+$.

\subsection{\label{sssec:X}$X^2\Pi_g$ state}

\begin{figure}[h]
\begin{minipage}[h]{\columnwidth}
\includegraphics[width=\textwidth]{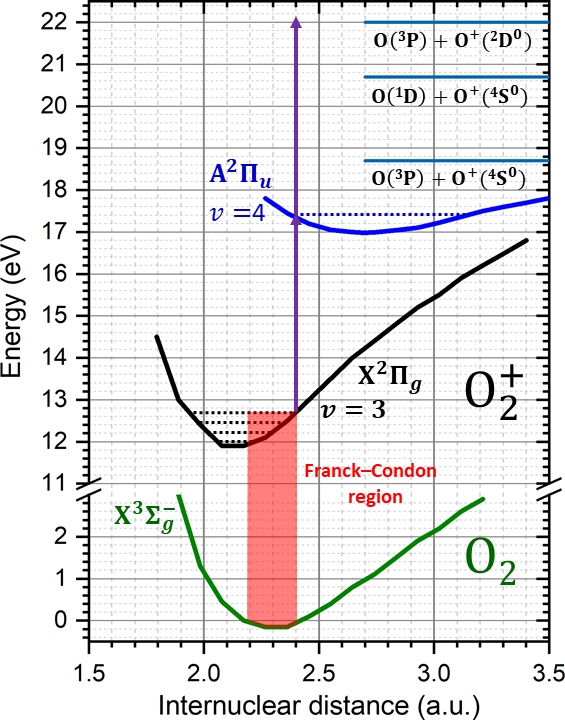}
\caption{Sketch of relevant potential energy curves for the dissociation of the $X^2\Pi_g$ state. The potential curves are adapted from Ref.~\cite{Krupenie1972}. The pump excites a rotational wave packet on O$_2^+$ $X^2\Pi_g$ state. This wave packet absorbs two UV probe photons and dissociates into different dissociation limits. The A-X transition is resonant with our probe wavelength at 264 nm. See Sec.~\ref{sssec:X} for more details.}
\label{fig:PES_X}
\end{minipage}
\end{figure}

In Fig.~\ref{fig:UV_probe}(a), the bound wave packet in the ground state $X^2\Pi_g$ of O$_2^+$ can be observed via two-photon dissociation into the first three dissociation limits: $L_1$ (near 3.2 eV KER), $L_2$ (near 1.4 eV KER) and $L_3$ (near zero KER). The vibrational level $v=3$ dominates all channels. Why $X^2\Pi_g (v=3)$ is dominant? According to the reference data provided by Krupenie (\cite{Krupenie1972}, page 472), there is a resonance between $X^2\Pi_g(v=3)$ and $A^2\Pi_u(v=4)$ of O$_2^+$ in the range of 263.27 nm to 264.67 nm (as described in Fig.~\ref{fig:PES_X}), and our probe spectrum is centered at 263.8 nm with a 2.4-nm bandwidth. The presence of this resonant process amplifies the contribution from $v=3$ relative to other vibrational levels in $X^2\Pi_g$ that may be populated by the pump.

\begin{figure*}
\begin{minipage}[b]{\textwidth}
\includegraphics[width=\textwidth]{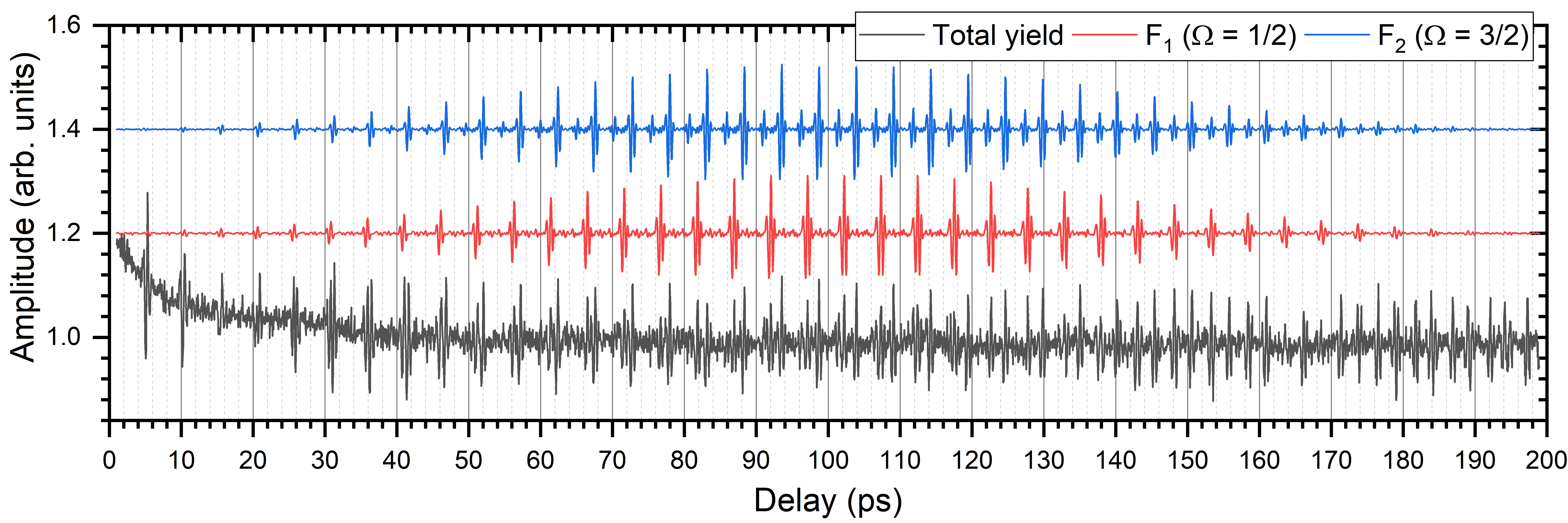}
\caption{Due to the spin-orbit coupling, the time-dependent yield of $X^2\Pi_g (v=3)$ dissociated into $L_2$ channel (integrating KER range from 1.2 eV to 1.7 eV) (black, bottom) is complicated and nonperiodic without any regular revival structures. Filtering in the frequency domain shows different time evolutions of F$_1$ and F$_2$ series. The $\Omega=3/2$ (blue, top) exhibits clear quarter revivals while it is not clear for the $\Omega=1/2$ (red, middle). The F$_1$ and F$_2$ curves are shifted up for clarity. The envelope of the $\mathrm{F_1}$ and $\mathrm{F_2}$ curves is the artifact from the Hanning window and the zero padding used in the FFT algorithm.}
\label{fig:FFT_separation}
\end{minipage}
\end{figure*}

For brevity, we will call this $A^2\Pi_u(v=4) \longleftarrow X^2\Pi_g(v=3)$ the 4--3 transition. The two nearby transitions are 2--2 (261.73--263.07~nm, very weak) and 6--4 (265.23--266.65~nm, weaker but comparable to 4--3 transition). Within the 2.4 nm bandwidth of the probe, there is some overlap with the 2--2 transition and very little overlap with the 6--4 transition. In the FFT of the ion yield, we observe a faint signature of $X^2\Pi_g(v=2)$, and no significant contribution of $X^2\Pi_g(v=4)$. Beside the lack of overlap with the spectrum of the UV probe, $X^2\Pi_g(v=4)$ is not efficiently populated by the pump since its wave function has less overlap with the neutral ground state wave function compared to lower vibrational levels (see Fig.~\ref{fig:PES_X}), which also explains the absence of $X^2\Pi_g(v=4)$.

From potential energy curves calculated by Liu \textit{et al.} \cite{Liu2015}, only the $1 ^2\Sigma_g^+$ can be the candidate for dissociation of $X^2\Pi_g$ into $L_1$. No dipole transitions can account for $L_2$ since only quartet states have this asymptote. Out of 36 states dissociating into $L_3$ [O$^+({}^2D^0)+$ O$({}^3P)$], eight states can contribute through dipole-allowed two-photon absorption to the dissociation of $X^2\Pi_g$ to $L_3$, including $B ^2\Sigma_g^-$, $(2,3)^2\Sigma_g^+$, $(2,3,4)^2\Pi_g$ and $(1,2)^2\Delta_g$. With the current data, we are unable to narrow down this list. Please refer to Table.~\ref{tab:b_asymptote} for a summary of states and their dissociation limits.

We take a closer look at $X^2\Pi_g (v=3)$ dissociated into $L_2$ (near 1.4 eV KER) since it is separated in energy from other channels. The time-dependent yield of this channel, shown in Fig.~\ref{fig:FFT_separation}, is complicated and nonperiodic.  This is a consequence of the coupling of rotational and electronic angular momenta. In Hund’s case (a) approximation, rotational  energies are split by spin-orbit interaction into two series called $F_1$ and $F_2$. When the spin-orbit coupling constant $A$ is positive, these correspond to $\Omega=1/2$ and $\Omega=3/2$, where $\Omega$ is the projection of the total angular momentum along the internuclear axis. Since the rotational angular momentum is always perpendicular to the internuclear axis in a diatomic molecule, $\Omega=\Lambda+\Sigma$ where $\Lambda$ and $\Sigma$ are projections of orbital and spin angular momenta on the internuclear axis. The eigenstates of the Hamiltonian are linear superpositions of Hund's case (a) states, with $F_1$ dominated by $\Omega=1/2$ and $F_2$ by $\Omega=3/2$.       

Indeed, as shown in Fig.~\ref{fig:UV_probe}(c), the FFT spectrum of the ion yield shows two series of lines, whose frequencies are in good agreement with the expected positions of Raman lines for purely rotational transitions within F$_1$ and F$_2$ levels. Most frequencies in this FFT spectrum correspond to $\Delta J = 2$; $\Delta J = 4$ beats are very weak and are not visible on this scale. In this case, we can filter out quantum beat frequencies corresponding to either F$_1$ or F$_2$ series and observe their time dependence separately. Fig.~\ref{fig:FFT_separation} shows that the time evolutions of these two series are quite different. The $\Omega=3/2$ exhibits clear quarter revivals while it is not clear for the $\Omega=1/2$. At early delays, F$_1$ and F$_2$ are almost in phase, leading to a fairly regular revival structure in the first $\approx$~30 ps. At longer time delays, F$_1$ and F$_2$ drift apart causing the complicated behavior of the delay-dependent yield.

It is worth noting that, in the case of $^2\Pi_g$ state, frequencies corresponding to each $\ket{J,\Omega}$ level are expected to be degenerate (see, for instance, \cite{brown_carrington_2003}). Approximately, this can be interpreted as a consequence of the $\Lambda$-doubling effect \cite{Mulliken1931} where the coupling between rotation with electronic angular momentum splits the $\Lambda \ne 0$ term into two nearby levels. However, for homonuclear diatomic molecules, only one of these two levels is allowed due to symmetry restriction. In our case of $^{16}\text{O}_2^+$, Bose-Einstein statistics requires the wave function of each level to be symmetric with respect to inversion through the center of the molecule since $^{16}\text{O}$ has a nuclear spin $I=0$ (see, for instance, Fig. 1 of \cite{Coxon1984}), hence, asymmetric states are absent. Therefore, our label of quantum beats in Fig.~\ref{fig:UV_probe}(c) is nondegenerate. This $\Lambda$-doubling effect is not observed in the $b^4\Sigma^{-}_g$ state of O$_2^+$ and the ground state ${}^3\Sigma_g^{-}$ of O$_2$ \cite{Wangjam2021} since $\Lambda=0$ for these states.

The couplings between spin and other motions are typically not observed in the ground states of neutral molecules since the total spin is usually zero. Examples where these couplings were observed in gas phase diatomic molecules are the doublet ground state of NO \cite{Zhang2015} and the triplet ground state of O$_2$ \cite{Wangjam2021}. However, these couplings are crucial in determining the dynamics of excited and ionic states of molecules with higher spin multiplicity \cite{Yaghlane2007}.

\section{CONCLUSIONS}
Our experiment shows clear evidence of the residual ionic wave packet in the $b^4\Sigma^-_g$ state, predominantly in the $\nu=3$ and $\nu=4$ levels with both IR and UV probe pulses. The similarity in appearance of the $b^4\Sigma^-_g$ state quantum beat spectra with two different probes confirms the resonant coupling between $b^4\Sigma^-_g$ and $a^4\Pi_u$ induced by the 800-nm pump. On the other hand, the dominance of the $X^2\Pi_g (v=3)$ state shows the importance of resonant coupling in the UV probe pulse. Our experiment also reveals the importance of rovibrational excitation and predissociation in determining the momentum spectrum of the O$^+$ fragments. This provides valuable insights into the laser-induced dissociation of O$_2^+$ \cite{De2011, Zohrabi2011, Magrakvelidze2012, Corlin2015, Malakar2018, Ma2019, Fukahori2020, Yu2020, Xue2022, Rebholz2022, fukahori2024} and strongly supports the three-state model by Xue \textit{et al.} \cite{Xue2018}. We observe a very broad rotational wave packet in the $b^4\Sigma^-_g$ state. We see weak signatures of a wave packet in the $a^4\Pi_u$ state. It can be owing to a strong population redistribution into the $b^4\Sigma^-_g$ state, or the rotational spectrum of the $a^4\Pi_u$ is too spread out.

The rotational spectra that we observe are strikingly clean and simple, even for a diatomic molecule. Long-scan FFT spectroscopy with a “weak” probe pulse does not expose to short-lived states and deeply bound long-lived states. It can expose to long-lived states within a photon or two of a dissociation limit (selection rules permitting). As the anharmonicity of the potential energy curves (or surfaces) increases, the spectra will be more complicated, making it more difficult to identify states. However, at least for diatomic molecules, we believe that high-resolution Fourier spectra obtained from this type of measurement (can be with different pump-probe wavelengths) will allow unambiguous identification of the electronic, vibrational and rotational states, providing valuable insights into the laser-induced dynamics of the molecular cations.

Another interesting system that could be further investigated using this technique is the strong-field ionization of N$_2$, which plays an important role in air lasing \cite{Luo2003}, exhibits rotational coherence (see, for instance, \cite{Zhang2013}), and its mechanism is still under debate \cite{Liu2015b,Yao2016,Ando2019,Richter2020,Kleine2022}.

\begin{acknowledgments}
We thank D. Rolles, A. S. Venkatachalam, C. Blaga, S. Hosseini-Zavareh, and E. Mullins for their help with the FLAME and HITS lasers. This work was supported by the Chemical Sciences, Geosciences and Biosciences Division, Office of Basic Energy Sciences, Office of Science, U.S. Department of Energy under Award No. DE-FG02-86ER13491.
\end{acknowledgments}

\appendix
\section{\label{sec:Dunham_expansion} Dunham expansion for calculating rotational energy levels}

For the $b^4\Sigma^-_g$ state, rotation energies $E_N$ depend on the vibration quantum number $\nu$ as
\begin{equation}
    E_{\nu,N}=B_\nu N(N+1) - D_\nu[N(N+1)]^2,
    \label{eq:E_N}
\end{equation}
where the rotational constant is
\begin{equation}
    B_\nu=B_e-\alpha_e \left( \nu+\frac{1}{2} \right),
    \label{eq:B}
\end{equation}
and the centrifugal distortion is
\begin{equation}
    D_\nu=D_e+\beta_e\left(\nu+\frac{1}{2}\right).
    \label{eq:D}
\end{equation}
$N$ is the nuclear angular momentum quantum number, $B_e$ is the rotational constant in equilibrium position, $\alpha_e$ is the first correction term of the rotational constant, $D_e$ is the centrifugal distortion constant, and $\beta_e$ is the first correction term of the centrifugal distortion.

The quartet nature of the $b^4\Sigma^-_g$ state can be ignored for our experimental resolution ($<$0.169~cm$^{-1}$) since the spin-spin and spin-rotation constants are small ($\le0.14$ cm$^{-1}$) \cite{Albritton1977}. These constants for $b^4\Sigma^-_g$ have been measured \cite{Krupenie1972, Albritton1977,Cosby1980,Hansen1981,Hansen1983}. Higher order corrections are ignored.

For the $X^2\Pi_g$ state, beside the rotational constant $B_\nu$ and the centrifugal distortion $D_\nu$, the spin-orbit coupling constant $A_\nu$ is included. The spin-rotation coupling constant and the $\Lambda$-doubling constants ($p_\nu$ and $q_\nu$) are small and not included. The mentioned constants are provided in table IV of \cite{Coxon1984}. The rotation energy levels of the $X^2\Pi_g$ state are then determined by
\begin{eqnarray}
E_{vJ}^{\mp}&=&B_\nu(J^2-1\pm0.5X_\nu)\\
&&-D_\nu J^2(J^2+1\pm(4J/X_\nu)(J-Y_\nu/2J)),\nonumber
\end{eqnarray}
where $Y_\nu=A_\nu/B_\nu$, $X_\nu=\sqrt{4J^2+Y_\nu(Y_\nu-4)}$ and $A_\nu$ is the spin-orbit coupling constant (see, for instance, \cite{brown_carrington_2003}). $E^-_{vJ}$ is energy level of F$_1(\Omega=1/2)$ series and $E^+_{vJ}$ is energy level of F$_2(\Omega=3/2)$ series.



\section{\label{sec:states_n_limits} Summary of states and their dissociation limits}

\begin{table}[]
\begin{tabular}{|lll|}
\hline
\multicolumn{1}{|c|}{\begin{tabular}[c]{@{}c@{}}Potential\\ candidate\end{tabular}} & \multicolumn{1}{c|}{\begin{tabular}[c]{@{}c@{}}Energy\\ unlikely\end{tabular}} & \multicolumn{1}{c|}{\begin{tabular}[c]{@{}c@{}}Symmetry\\ forbidden\end{tabular}} \\ \hline\hline
\multicolumn{3}{|c|}{$b^4\Sigma^{-}_g$ + 2 $\hbar \omega$(800 nm) $\longrightarrow L_2$}\\ \hline
\multicolumn{1}{|c|}{$1^4\Delta_g$} & \multicolumn{1}{c|}{${2}^4\Pi_g$} & \multicolumn{1}{c|}{${1}^4\Sigma_u^-$, ${2}^4\Pi_u$, $1^4\Delta_u$}\\ \hline\hline
\multicolumn{3}{|c|}{$b^4\Sigma^{-}_g$ + $\hbar \omega$(264 nm) $\longrightarrow L_1$} \\ \hline
\multicolumn{1}{|c|}{$a^4\Pi_u$} & \multicolumn{1}{c|}{} & \multicolumn{1}{c|}{${d}^4\Sigma_g^+$, ${1}^4\Sigma_u^+$, $f^4\Pi_g$}\\
\multicolumn{1}{|c|}{} & \multicolumn{1}{c|}{} & \multicolumn{1}{c|}{all 8 doublet and sextet states}\\ \hline\hline
\multicolumn{3}{|c|}{$b^4\Sigma^{-}_g$ + $\hbar \omega$(264 nm) $\longrightarrow L_2$}\\ \hline
\multicolumn{1}{|c|}{$c^4\Sigma^{-}_u$, $2^4\Pi_u$} & \multicolumn{1}{c|}{} & \multicolumn{1}{c|}{${2}^4\Pi_g$, $1^4\Delta_g$, $1^4\Delta_u$}\\ \hline\hline
\multicolumn{3}{|c|}{$X^2\Pi_g+2\hbar \omega$(264 nm) $\longrightarrow L_1$}\\ \hline
\multicolumn{1}{|c|}{$1 ^2\Sigma_g^+$} & \multicolumn{1}{c|}{} & \multicolumn{1}{c|}{$1 ^2\Sigma_u^+$, $1^2\Pi_u$}\\
\multicolumn{1}{|c|}{} & \multicolumn{1}{c|}{} & \multicolumn{1}{c|}{all 8 quartet and sextet states}\\ \hline\hline
\multicolumn{3}{|c|}{$X^2\Pi_g+2\hbar \omega$(264 nm) $\longrightarrow L_2$}\\ \hline\multicolumn{1}{|c|}{} & \multicolumn{1}{c|}{} & \multicolumn{1}{c|}{all 6 quartet states}\\ \hline\hline
\multicolumn{3}{|c|}{$X^2\Pi_g+2\hbar \omega$(264 nm) $\longrightarrow L_3$}\\ \hline
\multicolumn{1}{|c|}{$B ^2\Sigma_g^-$, $(2,3)^2\Sigma_g^+$} & \multicolumn{1}{c|}{} & \multicolumn{1}{c|}{$1^2\Sigma_u^-$, $(2,3)^2\Sigma_u^+$, $(2,3,4)^2\Pi_u$}\\
\multicolumn{1}{|c|}{ $(2,3,4)^2\Pi_g$} & \multicolumn{1}{c|}{} & \multicolumn{1}{c|}{$(1,2)^2\Delta_u$, $1^2\Phi_g$, $1^2\Phi_u$}\\
\multicolumn{1}{|c|}{$(1,2)^2\Delta_g$} & \multicolumn{1}{c|}{} & \multicolumn{1}{c|}{all 18 quartet states}\\\hline
\end{tabular}
\caption{Potential states that can dissociate O$_2^+$ $b^4\Sigma^{-}_g$ and O$_2^+$ $X^2\Pi_g$ states into the first three dissociation limits ($L_1,L_2,L_3$) following dipole selection rules.}
\label{tab:b_asymptote}
\end{table}

\section{\label{sec:QB_freq} Calculated and measured quantum beat frequencies for selected channels}
In this Appendix, we give the calculated and the measured rotational quantum beat frequencies for different dissociation channels presented in the paper. The frequencies are in cm$^{-1}$.

\begin{table}[h]
\begin{tabular}{|r|r|r|r|r|r|r|}
\hline
\multicolumn{7}{|c|}{$b^4\Sigma^-_g$} \\ \hline
 & \multicolumn{6}{|c|}{$\nu = 4$} \\ \hline
$N$ & \multicolumn{1}{|c|}{$\Delta N = 2$} & \multicolumn{1}{|c|}{Expt.} & \multicolumn{1}{|c|}{$\Delta N = 4$} & \multicolumn{1}{|c|}{Expt.} & \multicolumn{1}{|c|}{$\Delta N = 6$} & \multicolumn{1}{|c|}{Expt.}\\ \hline
1 & 11.87 & -- & 33.23 & -- & 64.08 & -- \\ \hline
3 & 21.36 & 21.30 & 52.21 & 52.17 & 92.53 & -- \\ \hline
5 & 30.85 & 30.81 & 71.17 & 71.16 & 120.97 & 120.90  \\ \hline
7 & 40.33 & 10.33 & 90.12 & 90.10 & 149.37 & 149.28 \\ \hline
9 & 49.79 & 49.77 & 109.04 & 108.98 & 177.73 & 177.65  \\ \hline
11 & 59.25 & 59.22 & 127.93 & 127.88 & 206.04 & --  \\ \hline
13 & 68.69 & 68.65 & 146.79 & -- & -- & --  \\ \hline
15 & 78.11 & 78.06 & 165.61 & -- & -- & --  \\ \hline
17 & 87.51 & 87.46 & 184.39 & -- & -- & --  \\ \hline
\multicolumn{1}{|r|}{19} & \multicolumn{1}{|r|}{96.88} & \multicolumn{1}{|r|}{96.83} & \multicolumn{1}{|r|}{203.11} & \multicolumn{1}{|r|}{--} & \multicolumn{2}{|c|}{$\nu = 3,\Delta N=2$}\\ \hline
21 & 106.23 & 106.18 & 221.79 & -- & 108.22 & 108.24  \\ \hline
23 & 115.55 & 115.50 & 240.40 & -- & 117.72 & 117.77  \\ \hline
25 & 124.84 & 124.79 & 258.95 & -- & 127.19 & 127.25 \\ \hline
27 & 134.10 & 134.01 & 277.43 & -- & 136.63 & 136.70  \\ \hline
29 & 152.51 & 152.57 & 295.83 & -- & 146.03 & 146.05  \\ \hline
31 &  &  &  & -- & 155.40 & 155.46  \\ \hline
33 &  &  &  & -- & 164.72 & 164.80  \\ \hline
\end{tabular}
\caption{Calculated and measured rotational quantum beats for the $b^4\Sigma^-_g$ state (KER is integrated over the range of 0 to 130 meV). This data is discussed in Sec.~\ref{sec:0_250meV}.}
\label{tab:b_800nm_lowE}
\end{table}

\begin{table}[h]
\begin{tabular}{|r|r|r|r|r|r|r|r|r|}
\hline
\multicolumn{9}{|c|}{$b^4\Sigma^-_g$} \\ \hline
 & \multicolumn{2}{|c|}{$\nu = 2$} & \multicolumn{2}{|c|}{$\nu = 3$} & \multicolumn{2}{|c|}{$\nu = 3$} & \multicolumn{2}{|c|}{$\nu = 4$}\\ 
 & \multicolumn{2}{|c|}{$\Delta N = 2$} & \multicolumn{2}{|c|}{$\Delta N = 2$} & \multicolumn{2}{|c|}{$\Delta N = 4$} & \multicolumn{2}{|c|}{$\Delta N = 2$}\\ \hline
$N$ & \multicolumn{1}{|c|}{Calc.} & \multicolumn{1}{|c|}{Expt.} & \multicolumn{1}{|c|}{Calc.} & \multicolumn{1}{|c|}{Expt.} & \multicolumn{1}{|c|}{Calc.} & \multicolumn{1}{|c|}{Expt.} & \multicolumn{1}{|c|}{Calc.} & \multicolumn{1}{|c|}{Expt.}\\ \hline
13 & 71.30 & -- & 69.96 & -- & 149.53 & 149.72 & 68.69 & --\\ \hline
15 & 81.09 & 81.32 & 79.56 & 79.64 & 168.70 & 168.67 & 78.11 & 78.13\\ \hline
17 & 90.85 & 90.84 & 89.14 & 89.20 & 187.83 & -- & 87.51 & 87.35\\ \hline
19 & 100.59 & 100.60 & 98.69 & 98.75 & 206.91 & -- & 96.88 & 96.91\\ \hline
21 & 110.31 & 110.32 & 108.22 & 108.31 & 225.94 & -- & 106.23 & 106.13 \\ \hline
23 & 120.00 & 120.04 & 117.72 & 117.70 & 244.91 & -- & 115.55 & 115.52 \\ \hline
25 & 129.66 & -- & 127.19 & 127.25 &  & -- & 124.84 & 124.74\\ \hline
27 & 139.28 & 139.33 & 136.63 & 136.64 &  & -- & 134.10 & 133.96\\ \hline
29 & 148.88 & 148.88 & 146.03 & 146.03 &  & -- & 143.32 & --\\ \hline
31 & 158.43 & 158.44 & 155.40 & 155.42 &  & -- & 152.51 & --\\ \hline
33 & 167.95 & 168.00 & 164.72 & 164.81 &  & -- & 161.65 & --\\ \hline
35 & 177.43 & 177.38 & 174.00 & 174.03 &  & -- & 170.75 & --\\ \hline
\end{tabular}
\caption{Calculated and measured rotational quantum beats for the $b^4\Sigma^{-}_g$ state dissociated into the second dissociation limit ($L_2$). The KER is integrated over the range of 1.0 to 1.4 eV. This data is discussed in Sec.~\ref{sec:750_1500meV}.}
\label{tab:b_800nm_twophoton}
\end{table}

\begin{table}[h]
\begin{tabular}{|r|r|r|r|r|r|r|r|r|}
\hline
\multicolumn{9}{|c|}{$b^4\Sigma^-_g,\Delta N = 2$} \\ \hline
 & \multicolumn{2}{|c|}{$\nu = 2$} & \multicolumn{2}{|c|}{$\nu = 3$} & \multicolumn{2}{|c|}{$\nu = 4$} & \multicolumn{2}{|c|}{$\nu = 5$}\\ \hline
$N$ & \multicolumn{1}{|c|}{Calc.} & \multicolumn{1}{|c|}{Expt.} & \multicolumn{1}{|c|}{Calc.} & \multicolumn{1}{|c|}{Expt.} & \multicolumn{1}{|c|}{Calc.} & \multicolumn{1}{|c|}{Expt.} & \multicolumn{1}{|c|}{Calc.} & \multicolumn{1}{|c|}{Expt.}\\ \hline
1 & 12.32 & -- & 12.09 & 12.04 & 11.87 & 11.81 & 11.63 & 11.59\\ \hline
3 & 22.17 & -- & 21.76 & 21.78 & 21.36 & 21.33 & 20.93 & 20.93\\ \hline
5 & 32.02 & -- & 31.42 & 31.43 & 30.85 & 30.79 & 30.22 & 30.17 \\ \hline
7 & 41.86 & -- & 41.07 & 41.09 & 40.33 & 40.31 & 39.51 & 39.50\\ \hline
9 & 51.69 & -- & 50.72 & 50.73 & 49.79 & 49.77 & 48.78 & 48.68\\ \hline
11 & 61.50 & 61.59 & 60.35 & 60.37 & 59.25 & 59.21 & 58.05 & 57.98\\ \hline
13 & 71.30 & 71.34 & 69.96 & 69.99 & 68.69 & 68.64 & 67.29 & 67.25\\ \hline
15 & 81.09 & 81.12 & 79.56 & 79.58 & 78.11 & 78.06 & 76.52 & -- \\ \hline
17 & 90.85 & 90.82 & 89.14 & 89.17 & 87.51 & 87.45 & 85.72 & -- \\ \hline
19 & 100.59 & -- & 98.69 & 98.73 & 96.88 & 96.86 & 94.91 & -- \\ \hline
\end{tabular}
\caption{Calculated and measured rotational quantum beats for the $b^4\Sigma^{-}_g$ state dissociated into the first dissociation limit ($L_1$). The KER is integrated over the range of 4.1 to 5.0 eV. This data is discussed in Sec.~\ref{sssec:b}.}
\label{tab:b_800nm_onephoton}
\end{table}

\begin{table}[h]
\begin{tabular}{|r|r|r|r|r|r|r|}
\hline
\multicolumn{5}{|c|}{$X^2\Pi_g,\nu=3$} \\ \hline
 & \multicolumn{2}{|c|}{$\Omega = 1/2,\Delta N = 2$} & \multicolumn{2}{|c|}{$\Omega = 3/2,\Delta N = 2$} \\ \hline
$J$ & \multicolumn{1}{|c|}{Calc.} & \multicolumn{1}{|c|}{Expt.} & \multicolumn{1}{|c|}{Calc.} & \multicolumn{1}{|c|}{Expt.}\\ \hline
1/2 & 13.08 & -- & &  \\ \hline
1 1/2 & 19.62 & -- & 19.30 & -- \\ \hline
2 1/2 & 26.16 & 26.14 & 25.73 & 25.70  \\ \hline
3 1/2 & 32.70 & 32.71 & 32.16 & 32.16  \\ \hline
4 1/2 & 39.24 & 39.25 & 38.59 & 38.57  \\ \hline
5 1/2 & 45.77 & 45.78 & 45.02 & 45.04  \\ \hline
6 1/2 & 52.30 & 52.31 & 51.45 & 51.42  \\ \hline
7 1/2 & 58.83 & 58.84 & 57.87 & 57.88  \\ \hline
8 1/2 & 65.36 & 65.35 & 64.30 & 64.29  \\ \hline
9 1/2 & 71.88 & 71.90 & 70.72 & 70.73  \\ \hline
10 1/2 & 78.40 & 78.34 & 77.14 & 77.13  \\ \hline
11 1/2 & 84.92 & 84.94 & 83.56 & 83.56  \\ \hline
12 1/2 & 91.43 & 91.48 & 89.97 & 89.99  \\ \hline
13 1/2 & 97.94 & 97.89 & 96.38 & 96.38  \\ \hline
14 1/2 &  & & 102.79 & 102.83  \\ \hline
\end{tabular}
\caption{Calculated and measured rotational quantum beats for the $X^2\Pi_g(\nu=3)$ state dissociating into $L_2$. The KER is integrated over the range of 1.1 to 1.7 eV. This data is discussed in Sec.~\ref{sssec:X}.}
\label{tab:X}
\end{table}

\clearpage
\bibliography{Oplus_FFT}
\end{document}